# Different types of attacks in Mobile ADHOC Network: Prevention and mitigation techniques

***Aniruddha Bhattacharyya***
aniruddha.aot@gmail.com

***Arnab Banerjee***
arnab.saheb.85@gmail.com

***Dipayan Bose***
eth.hck23@gmail.com

***Himadri Nath Saha***

***Debika Bhattacharjee***

Department of Computer Science & Engineering, Institute Of Engineering & Management, Saltlake

Abstract :

*Security in mobile ADHOC network is a big challenge as it has no centralized authority which can supervise the individual nodes operating in the network. The attacks can come from both inside the network and from the outside. We are trying to classify the existing attacks into two broad categories: DATA traffic attacks and CONTROL traffic attacks. We will also be discussing the presently proposed methods of mitigating those attacks.*

**Keywords**    Network security, Ad-Hoc network

## 1. INTRODUCTION

A mobile ad hoc network (MANET) is a self-configuring network of mobile nodes. It lacks any fixed infrastructure like access points or base stations. It lacks centralized administration and is connected by wireless links/cables. Wireless ad hoc network can be build up where there is no support of wireless access or wired backbone is not feasible. All network services of ad hoc network are configured and created on the fly. Thus it is obvious that with lack of infrastructural support and susceptible wireless link attacks, security in ad hoc network becomes inherent weakness. Nodes within nomadic environment with access to common radio link can easily participate to set up ad hoc infrastructure. But the secure communication among nodes requires the secure communication link to communicate. Before establishing secure communication, link the node should be capable enough to identify another node. As a result node needs to provide his/her identity as well as associated credentials to another node. However delivered identity and credentials need to be authenticated and protected so that authenticity and integrity of delivered identity and credentials cannot be questioned by receiver node. Every node wants to be sure that delivered identity and credentials to recipient nodes are not compromised. Therefore it is essential to provide security architecture to secure ad hoc networking.

We found that many of the presently existing attacks have some common features and have been categorized into different attacks based on their minor differences. So hereby we are trying to categorize them into two broad categories: DATA traffic attacks and CONTROL traffic attacks. This will help in future designing of security measures which will be able in mitigating those broad categories in one go.

## 2. CLASSIFICATION OF ATTACKS

As previously discussed, we have categorized the presently existing attacks into two broad categories: DATA traffic attacks and CONTROL traffic attacks. This classification is based on their common characteristics and attack goals. For example: Black-Hole attack drops packets every time, while Gray-Hole attack also drops packets but its action is based on two conditions: time or sender node. But from network point of view, both attacks drop packets and Gray-Hole attack can be considered as a Black-Hole attack when it starts dropping packets. So they can be categorized under a single category.

There are few attacks that have implications on both DATA & CONTROL traffic, so they cannot be classified into these categories easily.

So those attacks are left for future discussions.

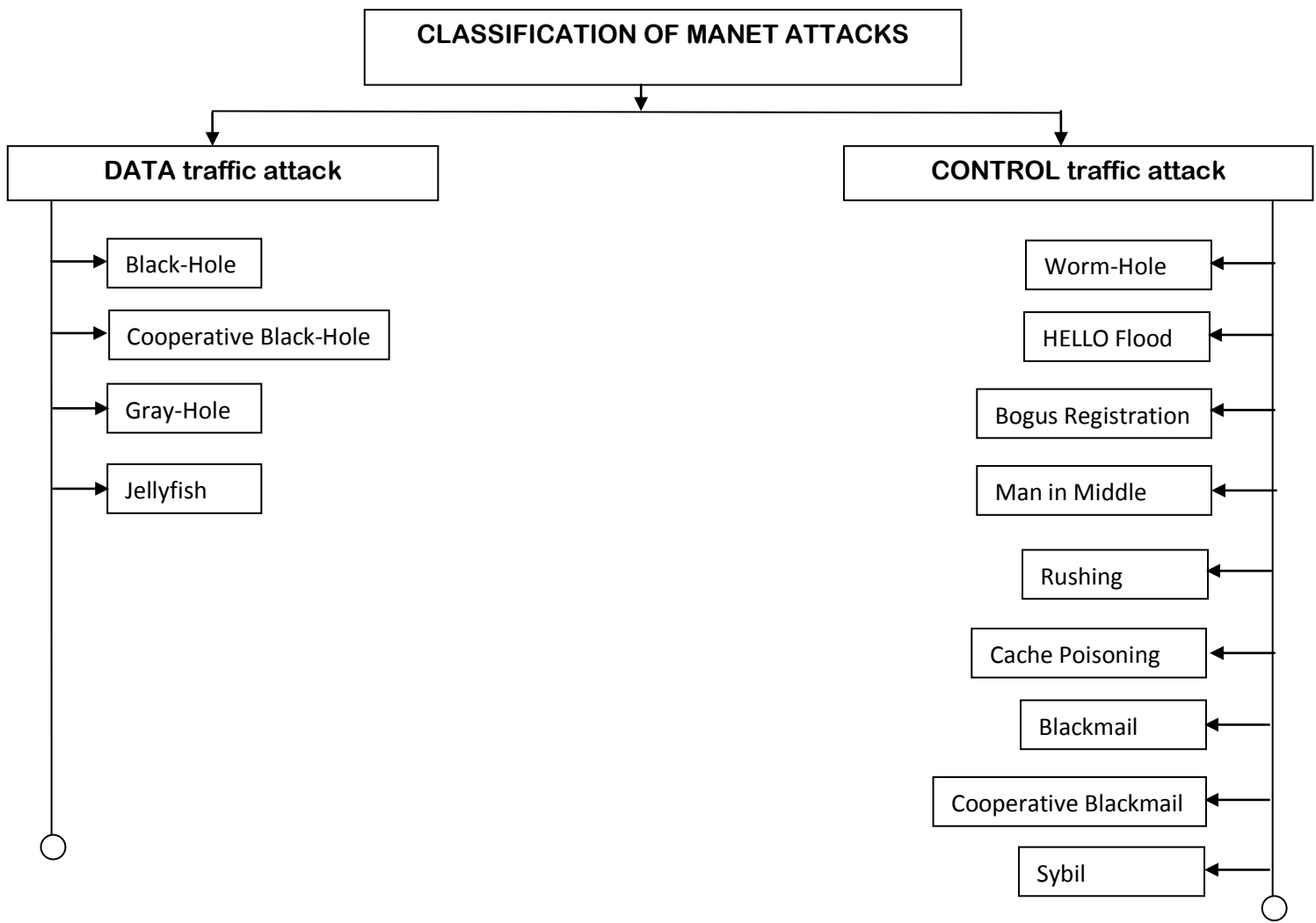

Figure 1: Classification of Mobile ADHOC Network (MANET) attacks

## 2.1 DATA Traffic Attack

DATA traffic attack deals either in nodes dropping data packets passing through them or in delaying of forwarding of the data packets. Some types of attacks choose victim packets for dropping while some of them drop all of them irrespective of sender nodes. This may highly degrade the quality of service and increases end to end delay. This also causes significant loss of important data. For e.g., a 100Mbps wireless link can behave as 1Mbps connection. Moreover, unless there is a redundant path around the erratic node, some of the nodes can be unreachable from each other altogether.

### 2.1.1 Black-Hole Attack [1][2][3][4]

In this attack, a malicious node acts like a Black hole, dropping all data packets passing through it as like matter and energy disappears from our universe in a black hole. If the attacking node is a connecting node of two connecting components of that network, then it effectively separates the network in to two disconnected components.

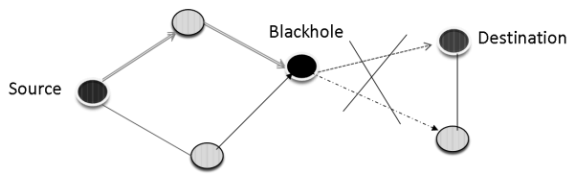

Figure 2: Black-Hole Attack

Here the Black-Hole node separates the network into two parts.

Few strategies to mitigate the problem:

(i) Collecting multiple RREP messages (from more than two nodes) and thus hoping multiple redundant paths to the destination node and then buffering the packets until a safe route is found.

(ii) Maintaining a table in each node with previous sequence number in increasing order. Each node before forwarding packets increases the sequence number. The sender node broadcasts RREQ to its neighbors and once this RREQ reaches the destination, it replies with a RREP with last packet sequence number. If the intermediate node finds that RREP contains a wrong sequence number, it understands that somewhere something went wrong.

### 2.1.2 Cooperative Black-Hole Attack [1][2][3]

This attack is similar to Black-Hole attack, but more than one malicious node tries to disrupt the network simultaneously. It is one of the most severe DATA traffic attack and can totally disrupt the operation of an Ad Hoc network. Mostly the only solution becomes finding alternating route to the destination, if at all exists.

Detection method is similar to ordinary Black-Hole attack.

In addition another solution is securing routing and node discovery in MANET by any suitable protocol such as SAODV, SNRP, SND, SRDP etc. Since each node is already trusted, black hole node should not be appearing in the network.

### 2.1.3 Gray-Hole Attack [9][10]

Gray-Hole attack has its own characteristic behavior. It too drops DATA packets, but node's malicious activity is limited to certain conditions or trigger. Two most common type of behavior:

(i) Node dependent attack – drops DATA packets destined towards a certain victim node or coming from certain node (fig 3), while for other nodes it behaves normally by routing DATA packets to the destination nodes correctly.

(ii) Time dependent attack – drops DATA packets based on some predetermined/trigger time while behaving normally during the other instances. (fig. 4)

Detecting this behaviorist attack is very difficult unless there exists a system wide detection algorithm, which takes care of all the nodes performance in the network. Sometimes nodes can interact with each other and can advise malicious nodes existence to other friendly nodes. Approach is similar to Black-Hole attack where sequence number feedback might detect some Gray-Hole attack. If multiple paths exist between sender and destination then buffering packets with proper acknowledgement (for e.g. 2ACK [14]) might detect active Gray-Hole attack in progress. But dormant or triggered attack is difficult to detect with this approach.

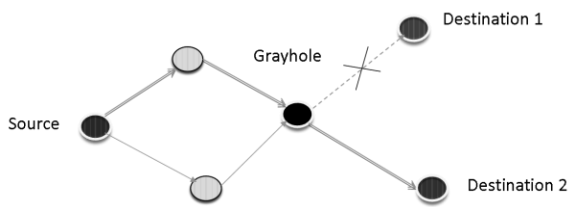

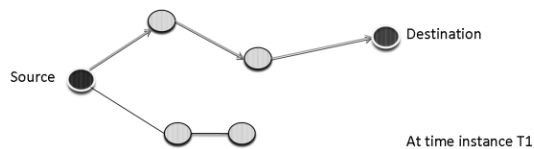

Figure 3: Gray-Hole – Node dependent attack

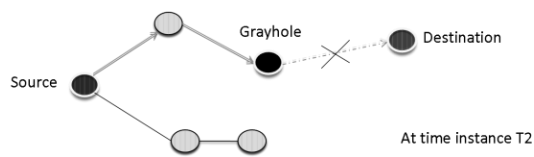

Figure 4: Gray-Hole – Time dependent attack

### 2.1.4 Jellyfish Attack [12][13][28]

Jellyfish attack is somewhat different from Black-Hole & Gray-Hole attack. Instead of blindly dropping the data packets, it delays them before finally delivering them. It may even scramble the order of packets in which they are received and sends it in random order. This disrupts the normal flow control mechanism used by nodes for reliable transmission. Jellyfish attack can result in significant end to end delay and thereby degrading QoS. Few of the methods used by attacker in this attack:

(i) One of the methods is scrambling packet order before finally delivering them instead of received FIFO order. ACK based flow control mechanism will generate duplicate ACK packets which will unnecessarily consume precious network bandwidth and battery life.

(ii) ) Another method can be, performing selective Black-Hole attack by dropping all packets at every RTO. This will cause timeout in sender node at every RTO for that duration. If nodes use traffic shaping, default flow control

mechanism might be triggered to the sender node as it is same as destination overwhelm

(iii) The attacking node can store all the received packets in its buffer but sends them after some random delay maintaining the received packet order. Here also the flow control mechanism gets confused. Sometimes the source node might take a longer route instead of the most obvious shortest route.

Few of the solutions to Jellyfish type attack includes:

(i) 2ACK [14] : The basic idea of the 2ACK scheme is that, when a node forwards a data packet successfully over the next hop, the destination node of the next-hop link will send back a special two-hop acknowledgment called 2ACK to indicate that the data packet has been received successfully. Such a 2ACK transmission takes place for only a fraction of data packets, but not for all.

(ii) Credit based systems [22]: This approach provides incentives for successful transmission of some kind of token or credit which the node might use when it starts sending its own packet.

(iii) Reputation based scheme [22]: Here individual nodes collectively detect misbehaving nodes (such as CONFIDANT).

## 2.2 CONTROL Traffic Attack

Mobile Ad-Hoc Network (MANET) is inherently vulnerable to attack due to its fundamental characteristics, such as open medium, distributed nodes, autonomy of nodes participation in network (nodes can join and leave the network on its will), lack of centralized authority which can enforce security on the network, distributed co-ordination and cooperation. The existing routing protocols can not be used in MANET due to these reasons.

Many of the routing protocols devised for use in MANET have their individual characteristic and rules. Two of the most widely used routing protocols is Ad-Hoc On Demand Distance Vector routing protocol (AODV), which relies on individual node's cooperation in establishing a valid routing table and Dynamic MANET On-Demand (DYMO) , which is a fast light weight routing protocol devised for multi hop networks. But each of them is based on trust on nodes participating in network. The first step in any successful attack requires the node to be part of that network. As there is no constraint in joining the network, malicious node can join and disrupts the network by hijacking the routing tables or bypassing valid routes. It can also eavesdrop on the network if the node can establish itself as the shortest route to any destination by exploiting the unsecure routing protocols. Therefore it is of utmost importance that the routing protocol should be as much secure as it can be.

Though there can be other kinds of attack, such as jamming attacks, which is not CONTROL attack. They can be tackled as a part of physical layer security protocols. Henceforth those attacks will not be discussed as are out of scope of this paper.

### 2.2.1 Worm Hole Attack [5][6][7]

Worm hole, in cosmological term, connects two distant points in space via a shortcut route. In the same way in MANET also one or more attacking node can disrupt routing by short-circuiting the network, thereby disrupting usual flow of packets. If this link becomes the lowest cost path to the destination then these malicious nodes will always be chosen while sending packets to that destination. The attacking node then can either monitor the traffic or can even disrupt the flow (via one of the DATA traffic attack). Wormhole attack can be done with single node also but generally two or more malicious node connects via a *wormhole-link*. In figure 5, Node X and Y performing wormhole attack.

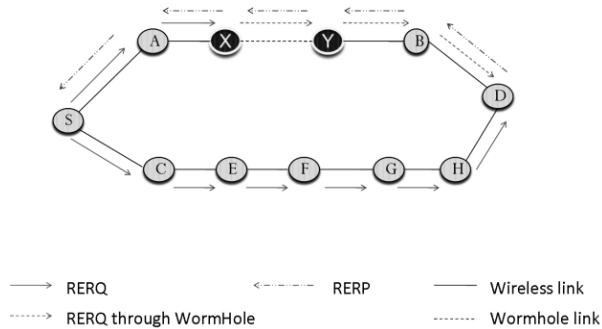

Figure 5: Worm-Hole attack

There have been few proposals recently to protect networks from worm-hole attack:

(i) Geographical leashes & temporal leashes: A leash is added to each packet in order to restrict the distance the packets are allowed to travel. A leash is associated with each hop. Thus, each transmission of a packet requires a new leash. A geographical leash is intended to limit the distance between the transmitter and the receiver of a packet. A temporal leash provides an upper bound on the lifetime of a packet.

(ii) Using directional antenna: Using directional antenna restricts the direction of signal propagation through air. This is one of the crude ways of limiting packet dispersion.

### 2.2.2 HELLO Flood Attack

The attacker node floods the network with a high quality route with a powerful transmitter. So, every node can forward their packets towards this node hoping it to be a better route to destination. Some can forward packets for those destinations which are out of the reach of the attacker node. A single high power transmitter can convince that all the nodes are his neighbor. The attacker node need not generate a legitimate traffic; it can just perform a selective replay attack as its power overwhelms other transceivers.

### 2.2.3 Bogus Registration Attack

A Bogus registration attack is an active attack in which an attacker disguises itself as another node either by sending stolen beacon or generating such false beacons to register himself with a node as a neighbor. Once registered, it can snoop transmitted packets or may disrupt the network altogether. But this type of attack is difficult to achieve as the attacker needs to intimately know the masquerading nodes identity and network topology. Encrypting packets before sending and secure authentication in route discovery (SRDP, SND, SNRP, ARAN, etc) will limit the severity of attack to some extent as attacker node has no previous knowledge of encryption method.

### 2.2.4 Man in Middle Attack [30]

In Man in Middle attack, the attacker node creeps into a valid route and tries to sniff packets flowing through it. To perform man in middle attack, the attacker first needs to be part of that route. It can do that by either temporarily disrupting the route by deregistering a node by sending malicious disassociation beacon captured previously or registering itself in next route timeout event. One way of protecting packets flowing through MANET from prying eyes is encrypting each packet. Though key distribution becomes a security issue.

### 2.2.5 Rushing Attack

In AODV or related protocol, each node before transmitting its data, first establishes a valid route to destination. Sender node broadcasts a RREQ (route request) message in neighborhood and valid routes replies with RREP (route reply) with proper route information. Some of the protocols use duplicate suppression mechanism to limit the route request and reply chatter in the network. Rushing attack exploits this duplicate suppression mechanism. Rushing attacker quickly forwards with a malicious RREP on behalf of some other node skipping any proper processing. Due to duplicate suppression, actual valid RREP message from valid node will be discarded and consequently the attacking node becomes part of the route. In rushing attack, attacker node does send packets to proper node after its own filtering is done, so from outside the network behaves normally as if nothing happened. But it might increase the delay in packet delivering to destination node.

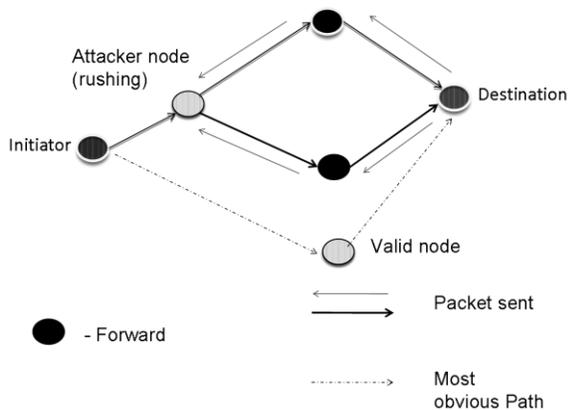

Figure 6: Rushing Attack

Few of the protocols that might help in resolving Rushing attack:

(i) SEDYMO [24]: Secured Dynamic MANET On-Demand is similar to DYMO but it dictates intermediate node must add routing information while broadcasting the routing messages and no intermediate node should delete any routing information from previous sender while broadcasting. It also incorporates hash chains and digital signature to protect the identity.

(ii) SRDP [23]: Secure Route Discovery Protocol is security enhanced Dynamic Source routing (DSR) protocol.

(iii) SND [26]: Secure Neighbor Detection is another method of verifying each neighbor's identity within a maximum transmission range.

### 2.2.6 Cache Poisoning Attack

Generally in AODV, each node keeps few of its most recent transmission routes until timeout occurs for each entry. So each route lingers for some time in node's memory. If some malicious node performs a routing attack then they will stay in node's route table until timeout occurs or a better route is found. An attacker node can advertise a zero metric to all of its destinations. Such route will not be overwritten unless timeout occurs. It can even advertise itself as a route to a distant node which is out of its reach. Once it becomes a part of the route, the attacker node can perform its malicious activity. Effect of Cache poisoning can be limited by either adding boundary leashes or by token authentication. Also each node can maintain its friend-foe list based on historical statistics of neighboring nodes performance.

Few of the mitigation methods proposed:

(i) SAODV [29]: Secure AODV is an extension to AODV protocol that adds each

node to exchange signed routing messages. Each node has its own public key which it uses to sign routing messages. Also SAODV uses hop count as a metric for shortest-route as AODV and uses hash chains to secure hop count information in route messages.

(ii) ARAN [16][18][28] : Authenticated Routing protocol for Ad-hoc Networks uses similar techniques as SAODV. ARAN uses certificates issued by a third party certification authority.

(iii) SNRP [16]: Secure Neighbor Routing protocol uses security enhanced Neighbor Lookup Protocol (NLP) to secure MANET routing. Newly added node uses public key to participate in MANET.

### 2.2.7 Blackmailing and Co-operative Blackmailing Attack

In a blackmailing attack or more effectively co-operative blackmailing attack, attacker nodes accuse an innocent node as harmful node. This attack can effectively be done on those distributed protocols that establish a good and bad node list based on review of participating nodes in MANET. Few of the protocols tries to make them more secure by using majority voting principle, but still if sufficient no. of attacker nodes become part of the MANET it can bypass that security also.

Another generic method of this attack will be, sending invalid RREP messages with advertising an unnecessarily high cost to certain nodes.

Known mitigation techniques:

(i) Dynamic Trust based, Distributed IDs [22]: As MANET routing is a co-operative process, while building a route each node must evaluate its neighbor nodes. This method builds a distributed trust relationships and maintain dynamic trust information. As the trust is part of a long chain, single malicious node cannot victimize an innocent node easily.

(ii) Friend List based [22]: Another solution will be building a friend list of trusted nodes. Nodes identity must be determined by the user who created the MANET. So it becomes a closed system of trusted nodes.

### 2.2.8 Sybil Attack

Sybil attack manifests itself by faking multiple identities by pretending to be consisting of multiple nodes in the network. So one single node can assume the role of multiple nodes and can monitor or hamper multiple nodes at a time. If Sybil attack is performed over a blackmailing attack, then level of disruption can be quite high. Success in Sybil attack depends on how the identities are generated in the system.

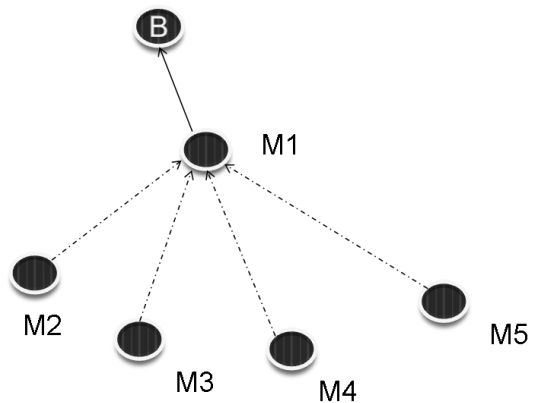

Figure 7: Sybil Attack

In figure 7, node M1 assumes identities of M2, M3, M4, and M5. So, to node B, M1 is equivalent to those nodes.

One way of mitigating this attack is maintaining a chain of trust, so single identity is generated by a hierarchical structure which may be hard to fake.

## 3. CONCLUSION

We have tried to categorize the different types of ad hoc security attacks solely based on their characteristics to considerably reduce the mitigation period. By bringing the attacks under these two broad categories the complicacy of naming also reduces. We have also kept a close look on the existing algorithms needed to mitigate the attacks and have tried to bind the attacks into categories according to that.

Some attacks have characteristics which makes them unsuitable to be categorized into these categories, so they have been kept away from this topic of discussion for the time being.

Further study is in progress to find out more common characteristics of the attacks to more strongly bind them into these categories and to ably design more powerful algorithm in mitigating DATA and CONTROL traffic attacks.